\documentclass{ws-ijmpd}
\usepackage[super]{cite}
\usepackage{xcolor}
\usepackage[verbose,hypertexnames=false]{hyperref}
\hypersetup{colorlinks=false,allbordercolors=blue,pdfborderstyle={/S/U/W 1}}
\newcommand{\al}{\alpha}
\newcommand{\g}{\gamma}

\newcommand{\s}{\sigma}

\newcommand{\e}{\mathrm{e}}
\newcommand{\be}{\begin{equation}}
\newcommand{\ee}{\end{equation}}

\newcommand{\f}[2]{\frac{#1}{#2}}

\definecolor{mygreen}{RGB}{82,180,78}
\definecolor{mypurple}{RGB}{126,70,225}
\definecolor{myblue}{RGB}{40,110,135}

\usepackage{tikz}
\usetikzlibrary{arrows.meta,positioning,calc}
\usepackage{tensor}
\usepackage{booktabs}

\usepackage{amsmath,amssymb,slashed,mathtools}
\usepackage{slashed}

\DeclareSymbolFontAlphabet{\amsmathbb}{AMSb}%

\usepackage{physics}
\usepackage{dsfont}

\begin{document}

\markboth{Alexia Nix}
{Wilson loops, M2-branes and strings}

%%%%%%%%%%%%%%%%%%%%% Publisher's Area please ignore %%%%%%%%%%%%%%%
%
\catchline{}{}{}{}{}
%
%%%%%%%%%%%%%%%%%%%%%%%%%%%%%%%%%%%%%%%%%%%%%%%%%%%%%%%%%%%%%%%%%%%%

\title{Wilson loops, M2-branes and strings}

\author{Alexia Nix}

\address{University of Iceland, Science Institute\\
Dunhaga 3, 107 Reykjav{\'i}k, Iceland\\
alexianix@hi.is}

\maketitle

%\begin{history}
%\received{(Day Month Year)}
%\revised{(Day Month Year)}
%\accepted{(Day Month Year)}
%\published{(Day Month Year)}
%\end{history}

\begin{abstract}
We establish a universal approach for studying the semiclassical quantization of both probe M2-branes and fundamental strings in asymptotically AdS$_{4}$ backgrounds. On the field theory side, 
their partition functions are dual to the vacuum expectation value of half-BPS Wilson loops of a class of three-dimensional $\mathcal{N}\geq 2$ Chern-Simons-matter theories. Utilizing this universal approach we arrive at novel predictions for the semiclassical behaviour of the Wilson loop or upon availability match our results with the previously obtained expressions via supersymmetric localization. We also show that the correct choice of holographic ensemble suggests that the semiclassical quantization of probe M2-branes yields the full perturbative answer in the rank of the gauge group $N$ for the dual Wilson loop. 
This is a proceedings contribution to the ``Athens Workshop in Theoretical Physics: 10th Anniversary'', held at the National and Kapodistrian University of Athens on December 17-19 2025.
\end{abstract}


\section{Introduction}

Recent advancements regarding the quantization of fundamental strings and  membranes in asymptotically AdS backgrounds in combination with results from supersymmetric localization have led to robust holographic tests of the AdS/CFT correspondence~\cite{Maldacena:1997re,Gubser:1998bc,Witten:1998qj} beyond the classical supergravity limit. 
In many cases, this has typically relied on having the field theory answer of the supersymmetric observable under consideration, to fine-tune the available tools needed to extract the corresponding gravity answer. A prime example of this, in the context of AdS$_{4}$/CFT$_{3}$, is the three-dimensional $\mathcal{N}=6$ U$(N)_{k}\times$U$(N)_{-k}$ Chern-Simons-matter theory known as ABJM theory, which is dual to AdS$_{4}\times S^{7}/\mathbf{Z}_{k}$ in M-theory or AdS$_{4}\times \mathbf{CP}^{3}$ in type IIA string theory~\cite{Aharony:2008ug}. The vacuum expectation value (vev) of the Wilson loop operator of this theory on $S^{3}$ was studied in the planar limit in Refs.~\citen{Kapustin:2009kz,Suyama:2009pd,Drukker:2009hy} using the tool of supersymmetric localization introduced in Ref.~\citen{Pestun:2007rz}, while Ref.~\citen{Marino:2009jd} used the resulting matrix model to make connections with topological string theory and extracted the exact expression for the Wilson loop in the planar limit.\footnote{In particular, Refs.~\citen{Kapustin:2009kz,Suyama:2009pd} studied the matrix model of the $1/6$-BPS Wilson loop, which was shown to be cohomologically equivalent to that of the $1/2$-BPS Wilson loop in Ref.~\citen{Drukker:2009hy}.} The first subleading planar correction for the Wilson loop was obtained in Ref.~\citen{Drukker:2010nc}, and the exact perturbative in $N$ answer for the same observable was derived in Ref.~\citen{Klemm:2012ii}, by using the Fermi gas approach. 

These field theory results paved the way for a number of holographic precision tests both from the string theory and the M-theory perspective, as the holographic dual of the Wilson loop in the fundamental representation is given by the partition function of a fundamental string (with the appropriate boundary conditions) in string theory~\cite{Rey:1998ik,Maldacena:1998im}, or the partition function of a probe M2-brane in M-theory. Specifically, the holographic Wilson loop of the ABJM theory has been studied from both perspectives. The supergravity result was matched up to one-loop order with the field theory answer in Refs.~\citen{Medina-Rincon:2019bcc,David:2019lhr}, by evaluating the ratio of two different supersymmetric holographic Wilson loops in AdS$_{4}\times \mathbf{CP}^{3}$, in order to cancel the dependence of the one-loop partition function on the divergences.  Subsequently, in Ref.~\citen{Giombi:2020mhz} a novel regularization scheme was introduced that allowed for the exact evaluation of the one-loop string partition function. From the M-theory perspective the same authors where able to recover the subleading behaviour of the half-BPS Wilson loop directly from the one-loop partition function of the membrane using $\zeta$-regularization \cite{Giombi:2023vzu}.

Although the tool of supersymmetric localization in combination with the Fermi gas approach provides an exact perturbative expression for the Wilson loop in ABJM theory, its model-specific nature limits its applicability to more general three-dimensional $\mathcal{N}\geq 2$ Chern–Simons–matter theories, where obtaining the full perturbative expansion of such observables remains challenging.\footnote{Some success in this direction has been achieved for a different observable, namely the free energy of $S^{3}$, for a specific class of three-dimensional $\mathcal{N}\geq 2$ Chern–Simons–matter theories. ~\cite{Marino:2011eh,Marino:2012az}}
Therefore, in this review based on Ref.~\citen{Gautason:2025bft}, instead of studying the field theory side to gain insights on gravity, we 
choose to use the previously mentioned successful tools to generate new results for the subleading or even perturbatively exact in $N$ answer for 
the half-BPS Wilson loop of two families of three-dimensional $\mathcal{N}\geq 2$ Chern–Simons–matter theories, directly from gravity. Family A contains superconformal field theories (SCFTs) whose gravity duals are given by the backreaction of $N$ coincident M2-branes that are located at the tip of a four-fold Calabi-Yau cone resulting in an AdS$_4\times$SE$_{7}$ eleven-dimensional supergravity background, where SE$_{7}$ denotes a seven-dimensional Sasaki-Einstein manifold. Evidently, the best studied example of this family is ABJM theory. The second family we study is coined family B and is composed of theories whose gravity duals take the form of a warped product of AdS$_{4}\times X_{6}$ in \textit{massive} type IIA supergravity where the internal space is asymptotically $\mathcal{S}(\text{SE}_{5})$, i.e. a sine-cone of a SE$_{5}$ manifold.~\cite{Guarino:2015jca,Fluder:2015eoa} Importantly, these backgrounds do not admit an M-theory uplift as the corresponding dilaton is bounded,~\cite{Aharony:2010af} and hence we will study the corresponding holographic Wilson loops from the string theory perspective.

For family A we study the holographic Wilson loops by building on the work of Ref.~\citen{Farquet:2013cwa}, where the importance of choosing the correct M-theory circle $S^{1}_{M}\subset \text{SE}_{7}$ was highlighted, and the classical membrane partition function was matched to the leading exponential behaviour of the vev of the half-BPS Wilson loop. In the case of family B, the successful holographic test that was performed in Ref.~\citen{Fluder:2015eoa}, and obtained a universal expression for the Wilson loop vev at leading order in the holographic limit, motivated us to explore whether this universality also persists beyond leading order. 
Remarkably, utilizing the geometric properties of these backgrounds, in Ref.~\citen{Gautason:2025bft} we show that both families admit a unified treatment for the semiclassical quantization of the fundamental string/M2-brane, which yields a universal expression for the corresponding one-loop partition function.

\section{Holographic Wilson loops in M-theory}

We start by reviewing the answer for the vev of the half-BPS Wilson loops of family A, whose field theories are three-dimensional U$(N)_{k_{1}}\times$U$(N)_{k_{2}}\times\dots \times$U$(N)_{k_{\mathcal{G}}}$ Chern-Simons-matter theories with $\sum_{i=1}^{\mathcal{G}}k_{i}=0$, where $k_{i} \in \mathbf{Z}$ is the Chern-Simons (CS) level associated to each U$(N)$ gauge group. For these theories the 't Hooft coupling is given by $\lambda=N/k$, where $k=\text{gcd}\{k_{i}\}$. In the large $N$ limit the corresponding matrix model for these theories has been studied in Refs.~\citen{Kapustin:2009kz,Hama:2010av, Herzog:2010hf} and led to the evaluation of the circular Wilson loop vev on $S^{3}$ to  leading order in Ref.~\citen{Farquet:2013cwa}
\begin{equation}\label{WL fam A LO}
	\log \langle W\rangle_{\text{A}}=2\pi^{3}c\sqrt{\frac{N}{6\text{vol}(\text{SE}_{7})}}\,,
\end{equation}  
where $c$ is inversely proportional to $k$, as it denotes the radius of the M-theory circle $S^{1}_{M}$. Although \eqref{WL fam A LO} is all that is currently known for the full set of family A, we should note that the Wilson loop vev of some theories such as the one of ABJM and ADHM theory has a known exact perturbative expression in $N$ and is proportional to a ratio of Airy functions.~\cite{Klemm:2012ii,Okuyama:2016pwb}

The holographic duals to family A are given by the AdS$_4\times$SE$_{7}$ eleven-dimensional supergravity background supported by a non-trivial four-form flux $G_{4}$ which reads
\begin{gather}
	\begin{aligned}
	\dd{s}^{2}_{11}&=\frac{R^{2}}{4}\dd{s}_{\text{AdS}_{4}}^{2}+R^{2}\dd{s}_{\text{SE}_{7}}^{2}\,,\\
	\dd{s}_{\text{SE}_{7}}^{2}&=\eta^{2}+\dd{s}_{\text{KE}_{6}}^{2}\,,\\
	G_{4}&=\frac{3}{8}R^{3}\text{vol}_{\text{AdS}_{4}}\,,
		\end{aligned}
\end{gather}
where $\dd{s}_{\text{SE}_{7}}^2$ is the metric on the Sasaki-Einstein manifold, and its contact one-form $\eta$ is related to the Kähler two-form $\omega$ of the Kähler-Einstein base manifold through $\dd{\eta}=2\omega$. From here it is evident that the supergravity approximation is valid when $R\gg \ell_p$, where $\ell_p$ is the eleven-dimensional Planck length. Furthermore, by relating the radius $R$ through the flux quantization condition 
\begin{equation}\label{flux quant}
	N=-\frac{1}{(2\pi\ell_{p})^{6}}\int \star_{11} G_{4}\,,
\end{equation}
with the field theory parameters we find
\begin{equation}
	\Big(\frac{R}{l_{p}}\Big)^{6}=\frac{(2\pi)^{6}kN}{6\text{vol}(\text{SE}_{7})}\,.
\end{equation}

The holographic Wilson loop of family A is given by the partition function of a probe M2-brane in the above solution, whose dimensionless tension is proportional to  $T_{\text{M2}}\sim\sqrt{k N}$. 
The partition function is then expanded around the large tension, i.e. large $N$ limit, as follows
\begin{equation}\label{WL expansion}
	\langle W \rangle=Z_{\text{M2}} \approx \e^{-\hat{S}_\text{cl}}\hat{Z}_{1-\text{loop}}\,,
\end{equation}
where $\hat{S}_\text{cl}$ is the leading classical contribution that was matched with \eqref{WL fam A LO} in Ref.~\citen{Farquet:2013cwa} and corresponds to the membrane action on its classical configuration.
The classical bosonic M2-brane action in static gauge reads~\cite{Bergshoeff:1987cm}
\begin{equation}\label{class M2}
	\hat{S}_{\text{cl}}=\frac{1}{(2\pi)^{2}\ell_p^{3}}\int \text{vol}_{3}-\frac{i}{(2\pi)^{2}\ell_p^{3}}\int A_{3}\,,
\end{equation}
where the induced volume form over the worldvolume of the M2-brane is given by $\text{vol}_{3}$ and $A_{3}$ denotes the pull-back of the three-form to the worldvolume.
The one-loop fluctuations around the classical configuration of the membrane are captured by the one-loop partition function $\hat Z_{1-\text{loop}}$ which is our object of interest.

In order to study these fluctuations it is important to understand the classical configuration of the membrane. This is given by a membrane wrapping the M-theory circle $S^{1}_{M}\subset \text{SE}_{7}$ and has an AdS$_{2}\times S^{1}_{M}$ worldvolume.  In order to determine the correct M-theory circle, one needs to introduce the Killing vector $\zeta_{M}$ which generates this U$(1)_{M}$ circle, and crucially is \textit{not} the Reeb vector $\zeta_{R}$ which satisfies $\zeta_{R}\lrcorner \dd{\eta}=0$ and $\zeta_{R}\lrcorner \eta=1$. However, it was shown in Ref.~\citen{Farquet:2013cwa} that supersymmetry requires the membrane to be located at points $p$ in the internal manifold at which $\zeta_{M}\lvert_{p}\propto\zeta_{R}$. This can be shown explicitly by defining a so-called Hamiltonian function $h_{M}$ as $h_{M}\coloneqq \zeta_{M}\lrcorner \eta$, and using the fact that the classical configuration of the BPS M2-brane is given by points $p$ for which this function is extremized, leading to~\cite{Farquet:2013cwa} 
\begin{equation}\label{zeta reeb}
	\zeta_{M}\lrcorner\dd{\eta}\lvert_{p}=0\,.
\end{equation}   

We are now ready to write down the induced metric of the membrane
\begin{equation}
\dd{s}^{2}_{\text{M2}}=\frac{R^{2}}{4}\dd{s}_{\text{AdS}_{2}}^{2}+R^{2}c^{2}\dd{\psi}^{2}\,,
\end{equation} 
where the on-shell radius of $S^{1}_{M}$ is given by $Rc$ with $c=h_{M}(p)$ and the $\psi$ coordinate is $2\pi$-periodic. For the above configuration of the M2-brane we find that the latter term of \eqref{class M2} vanishes and the classical action reduces to its worldvolume area 
\begin{equation}
	\hat{S}_{\text{cl}}=-2 c \mu\,, \qquad \mu\coloneqq R^{3}/(8\ell_p^{3})\,.
\end{equation}
As initially shown in Ref.~\citen{Farquet:2013cwa} this successfully matches the field theory answer and for future convenience we introduced $\mu$, which is the chemical potential on the field theory side. 

We now proceed to the semiclassical quantization of the M2-brane under consideration. To this end, we need to study the quadratic fluctuations of the normal bundle directions to the worldvolume around the classical configuration, which are made up of eight real scalars $\zeta^{a}$, with $a=1,\dots,8$ and eight fermions. The resulting action reads~\cite{Astesiano:2024sgi}

\begin{equation}\label{quad action M2}
	\begin{split}
		\hat{S}^{(2)}&=\frac{1}{(2\pi)^{2}\ell_{p}^{3}}\int\text{vol}_{3}\big( \mathcal{L}_{\text{bos}} + \mathcal{L}_{\text{ferm}} \big)\,,\\
		\mathcal{L}_{\rm{ferm}}&=2i\bar{\theta}\Big(\Gamma^i\partial_i+\frac{1}{4}\Gamma^{i}\tensor{\hat\Omega}{_{i}^{AB}}\Gamma_{AB}+\frac{1}{8}\slashed{G}-\f18 \Gamma^i \slashed{G}\Gamma_i\Big)\theta\,,\\
		\mathcal{L}_{\text{bos}}&=-\frac{1}{2}(\tensor{\hat{D}}{^{a}_{b}}\zeta^{b})^{2}+\frac{1}{2}(\tensor{\hat{R}}{^{i}_{aib}}+\tensor{\hat{K}}{_{a}^{ij}}\tensor{\hat{K}}{_{bij}}-\frac{i}{3!}\epsilon^{ijk}\nabla_{a}G_{bijk}+A_{Gica}\tensor{A}{_G^{ic}_b})\zeta^{a}\zeta^{b}\,,
	\end{split}
\end{equation}
with 
\begin{gather}\label{quad action M2 p2}
	\tensor{\hat{D}}{^{a}_{b}}\zeta^{b}=\nabla \zeta^{a}+\tensor{A}{^a_b}\zeta^{b}\,,\quad  \tensor{A}{_i^{ab}}=\tensor{\hat\Omega}{_{i}^{ab}}+\tensor{A}{_{Gi}^{ab}}\,,\quad \tensor{A}{_{Gi}^{ab}}=-\frac{i}{4}\epsilon_{ijk}G^{abjk}\,.
\end{gather}
Here we should impose the $\kappa$-symmetry gauge condition $\Gamma_{(3)}\theta = i\theta$ on the eleven-dimensional spinors $\theta$, where $\Gamma_{(3)}=\f1{3!}\epsilon^{ijk}\Gamma_{ijk}$ and $\epsilon^{ijk}$ is the three-dimensional Levi-Civita symbol. 
Moreover, the eleven-dimensional spin connection is given by $\tensor{\hat\Omega}{_{i}^{ab}}$, $\tensor{\hat R}{^{i}_{ajb}}$ is the Riemann tensor, $\hat K_{aij}$ is the extrinsic curvature, and  $\Gamma_{i}$ denote the eleven-dimensional gamma matrices that are pulled-back to the worldvolume of the M2-brane. 

In practice, we can carefully work out the non-trivial components of each object by using two important facts. The first is that due to \eqref{zeta reeb}, the tangent vector to the M-theory circle direction is proportional to the Reeb vector, which yields a universal expression for the three-dimensional mass term in the above action. Ultimately, this is due to the geometric properties of the SE$_{7}$ manifold. Second, since the spin connection does not benefit from this fact we need to evaluate its contribution with more care. It follows that when evaluated on-shell, the only components of the spin connection that have non-trivial value are the ones along SE$_{7}$. This means that the action only receives contributions from $\tensor{\hat\Omega}{_{\psi}^{ab}}$, where the $a,b$ components are taken along the KE$_{6}$ manifold. Then, we can parametrize this object in terms of its eigenvalues $\pm i q_{l}$, where $l=1,2,3$, which are related to the gauge charges of the fluctuations along the KE$_{6}$ directions and satisfy
\begin{equation}
	\sum_{l=1}^{3}q_{l}=-1\,,
\end{equation}
in accordance with the $\kappa$-symmetry gauge condition.  

Upon integration by parts we can rewrite the quadratic action in terms of the kinetic operators of the scalars $\hat{\cal K}_{ab}$ and fermions $\hat{\cal D}_{ab}$ on the three-dimensional worldvolume as
\begin{equation}\label{M2 quad action}
	\hat{S}^{(2)}= \frac{1}{(2\pi)^{2}\ell_{p}^{3}}\int\text{vol}_{3}(\zeta^{a}\hat{\cal K}_{ab}\zeta^{b}+\bar\theta^{a}\hat{\cal D}_{ab}\theta^{b})\,.
\end{equation}
The operators are diagonal in the $a,b$ indices and take a universal form that holds for any SE$_{7}$ manifold. This results in the following one-loop partition function 
\begin{equation}
\hat{Z}_{1-\rm{loop}}\coloneqq \e^{-\Gamma_{\text{M2}}}=\sqrt{\frac{ \prod_{\text{f}}\det \mathcal{\hat D}}{ \prod_{\text{b}}\det \mathcal{\hat K}}}\,,
\end{equation}
where a product is taken over all fermionic and scalar functional determinants, and we defined the effective M2-brane action $\Gamma_{\text{M2}}$ for future convenience. In order to reduce the problem to two-dimensions, we perform a Fourier mode expansion along the M-theory circle direction, which for the scalars reduces to~\cite{Giombi:2023vzu}
 \begin{equation}
 	\zeta^{a}=\sum_{n\in \mathbf{Z}} \e^{in\psi}\zeta_{n}^{a}(\sigma,\tau)\,,
 \end{equation}
 where $\sigma, \tau$ are the worldvolume coordinates along AdS$_{2}$, and a similar expression holds for the fermions. We proceed to evaluate the resulting effective action using the Heat kernel on AdS$_{2}$ together with $\zeta$-regularization,\footnote{Here the UV divergences arising from the evaluation of the functional determinants vanish due to $\zeta$-regularization.} which after some algebra yields~\cite{Gautason:2025bft} 
 \begin{gather}\label{eq: M2 eff action sum}
 		\Gamma_\text{M2}= -\sum_{n=-\infty}^\infty \sum_{l=0}^{3}\boldsymbol{\Gamma}^{(q_l,n)}\,,
 \end{gather}
 with
 \begin{equation}\label{eq: M2 eff action 2}
 \begin{split}
 	\boldsymbol{\Gamma}^{(q, n)} = &|1-q+ c^{-1}n |\bigg[\frac{1}{2}+\log\Gamma\Big(\frac{\abs{1-q+ c^{-1}n}}{2}\Big)\bigg]\\
 	&  -| q + c^{-1}n|\log\Gamma\Big(\f{1+\abs{q+ c^{-1}n}}{2}\Big)-\frac{1}{4}(1+2\log 2\pi)\\
 	&+2\psi^{(-2)}\Big(\f{1+\abs{q+ c^{-1}n}}{2}\Big)-2\psi^{(-2)}\Big(\f{\abs{1-q + c^{-1}n }}{2}\Big)\,,
 \end{split}
 \end{equation}
 where we defined $q_{0}\coloneqq3$. The resulting expression is entirely specified in terms of the charges $q_{l}$ that accompany a given SE$_{7}$ manifold and its corresponding M-theory radius $c$, but looks highly non-trivial. However, upon restricting to a given SE$_{7}$ manifold and inserting the corresponding quantities into the above expression, the effective action becomes remarkably simple and with the use of Euler's infinite product formula 
 \begin{equation}
 	\log \frac{\sin \pi x}{\pi x}= \sum_{n=1}^{\infty}\log(1-\frac{x^{2}}{n^{2}})\,,
 \end{equation}
often reduces to a trigonometric function.

 \subsection{An Airy proposal}
 
 Recent studies regarding the semiclassical quantization of M2-branes in holography have suggested that the resulting observables should not be compared directly with their field theory counterparts. The reason for this is that the two quantities are evaluated in different ensembles.~\cite{Gautason:2025plx} In particular, field theory observables are computed in the fixed $N$, i.e. canonical ensemble, whereas their holographic M2-brane counterparts are evaluated in the grand canonical ensemble, where instead the three-form potential $A_{3}$ i.e. $\mu$ is fixed.\footnote{See also Ref.~\citen{Bobev:2026gir} where a number of holographic precision tests were performed in the grand canonical ensemble.} Their M5-brane duals on the other hand are proposed to be evaluated in the canonical ensemble, as their semiclassical description requires knowing the $A_{6}$ potential. As the flux quantization condition \eqref{flux quant} suggests, this would indeed reduce to the fixed $N$ ensemble. However, the semiclassical quantization of M5-branes is currently out of reach. Therefore, we resort to  translating the M2-brane partition function to the canonical ensemble. To achieve this, it is not enough to simply relate $\mu$ with $N$ through their classical relation
 \begin{equation}
 	\mu^{2}=\frac{\pi^{6}N}{6\text{vol}(\text{SE}_{7})}\,,
 \end{equation}
 if we want to compare observables beyond the classical supergravity limit.
Instead, we must perform the following Laplace transform
 \begin{equation}\label{laplatrafo}
 \langle W(N) \rangle = \frac{1}{Z(N)}\frac{1}{2\pi i}\int_{C} \dd \mu \, \e^{\mathcal{Z}_{\rm{M2}}(\mu) - \mu N}\langle W(\mu) \rangle\,,
 \end{equation}
 with 
  \begin{equation}\label{laplatrafo2}
 Z(N) = \frac{1}{2\pi i}\int_{C} \dd \mu \, \e^{\mathcal{Z}_{\rm{M2}}(\mu) - \mu N}\,.
 \end{equation}
 Here $\mathcal{Z}_{\rm{M2}}(\mu)$ was shown in Ref.~\citen{Gautason:2025plx} to be given by the grand canonical potential $J(\mu)$ whose explicit form can be deduced from the Fermi gas analysis on the field theory side, and reads~\cite{Marino:2011eh,Marino:2012az}
 \begin{equation}
 	\mathcal{Z}_{\rm{M2}}(\mu)=J(\mu)=\frac{\mathcal{C}}{3}\mu^{3}+\mathcal{B}\mu+\mathcal{A}\,,
 \end{equation}
 where we neglected the non-perturbative corrections in $\mu$ as these would only contribute to the non-perturbative behaviour of the Wilson loop, which is beyond the scope of this review. Here $\mathcal{A},\mathcal{B},\mathcal{C}$ are independent of $N$ but can depend on other field theory parameters, and are in general fixed through a case-by-case analysis of the matrix model of each field theory.

From the perspective of the fixed $\mu$ ensemble many observables, including the Wilson loop vev of ABJM theory, were shown to be one-loop exact.~\cite{Gautason:2025plx} Therefore, we can assume that the same holds for our result for family A\footnote{From now on we suppress possible non-perturbative corrections to the Wilson loop vev.}
 \begin{equation}\label{WL famA}
 	\langle W(\mu) \rangle_{\text{A}}= \hat{Z}_{1-\rm{loop}}(k)e^{2 c \mu}\,,
 \end{equation}
 where the effective action of \eqref{eq: M2 eff action 2} only depends on the CS level $k$, through the M-theory radius $c$.
 Combining this with the Laplace transform \eqref{laplatrafo}, we find that the exact perturbative expression for the half-BPS Wilson loop vev of family A is given by~\cite{Gautason:2025bft}
 \begin{equation}\label{WL pert}
 	\langle W(N)\rangle_{\text{A}}=\hat{Z}_{1-\rm{loop}}(k)\frac{\text{Ai}(\mathcal{C}^{-1/3}(N-2c-\mathcal{B}))}{\text{Ai}(\mathcal{C}^{-1/3}(N-\mathcal{B}))}\,.
 \end{equation}

\subsection{Selected examples of Family A}
To demonstrate the predictive power of our above result, we first apply it to the ABJM and ADHM theories, for which the exact perturbative expression for the Wilson loop is known. As mentioned earlier, the ABJM theory is holographically dual to M-theory on AdS$_{4}\times S^{7}/\mathbf{Z}_{k}$. Using the fact that $q_{l}=\{3,1,-1,-1\}$ and the M-theory radius is $c=1/k$ in \eqref{eq: M2 eff action sum} and \eqref{eq: M2 eff action 2}, we find that the perturbative part of the Wilson loop of ABJM is given by 
\begin{gather}\label{WL pert ABJM}
	\langle W(N)\rangle_{\text{ABJM}}=\frac{1}{2\sin(\frac{2\pi}{k})}\frac{\text{Ai}\Big(\tfrac{\pi^{2/3}k^{1/3}}{2^{1/3}}(N-\tfrac{7}{3k}-\tfrac{k}{24})\Big)}{\text{Ai}\Big(\tfrac{\pi^{2/3}k^{1/3}}{2^{1/3}}(N-\tfrac{1}{3k}-\tfrac{k}{24})\Big)}\,,\qquad k>2\,,
\end{gather} 
where we inserted the field theory parameters $\mathcal{B},\mathcal{C}$ derived in Refs.~\citen{Marino:2011eh,Fuji:2011km}. As anticipated, this expression matches the respective field theory answer obtained in Ref.~\citen{Klemm:2012ii}.\footnote{As mentioned in the introduction, this holographic match was previously also achieved at subleading order in the large $N$ expansion in Ref.~\citen{Giombi:2023vzu}.}

The ADHM theory is a three-dimensional $\mathcal{N}=4$ U$(N)$ theory with vanishing Chern-Simons term, coupled to $N_{f}$ fundamental hypermultiplets and a single adjoint hypermultiplet. This theory describes the low energy dynamics of $N$ M2 branes probing $\mathbf{C}\times \mathbf{C}/\mathbf{Z}_{N_{f}}$ with an AdS$_{4}\times S^{7}/\mathbf{Z}_{N_{f}}$ near-horizon geometry.~\cite{Benini:2009qs} In this case, the charges are given by $q_{l}=\{3,-1,0,0\}$ and the M-theory radius is $c=1/N_{f}$. Using this together with \eqref{eq: M2 eff action sum}, \eqref{eq: M2 eff action 2} and \eqref{WL pert}, yields the following expression for the perturbative part of the Wilson loop vev of this theory
\begin{gather}\label{WL pert ADHM}
	\langle W(N)\rangle_{\text{ADHM}}=\frac{1}{4\sin(\frac{2\pi}{N_{f}})}\frac{\text{Ai}\Big(\tfrac{\pi^{2/3}N_{f}^{1/3}}{2^{1/3}}(N-\tfrac{5}{2N_{f}}+\tfrac{N_{f}}{8})\Big)}{\text{Ai}\Big(\tfrac{\pi^{2/3}N_{f}^{1/3}}{2^{1/3}}(N-\tfrac{1}{2N_{f}}+\tfrac{N_{f}}{8})\Big)}\,, \qquad N_{f}>2\,,
\end{gather} 
where we used the field theory data of Refs.~\citen{Jafferis:2011zi,Mezei:2013gqa,Grassi:2014vwa}. This result precisely matches the corresponding field theory answer of Ref.~\citen{Okuyama:2016pwb}.

We can then continue by evaluating  our expressions  for several examples of SE$_7$ manifolds that have not yet been analysed on the field theory side. This leads to new predictions for the perturbative Wilson loop vev for an infinite family of three-dimensional SCFTs.
A complete section in Ref.~\citen{Gautason:2025bft} is devoted to studying such examples. To avoid any extraneous material, we remark that for some examples we studied, the one-loop M2-brane partition function was found to equal the one-loop string partition function.
That is to say that it did not reduce to a trigonometric\footnote{As was the case for the above examples.} or more complicated function of the CS level $k$, but was simply given by $\hat{Z}_{1-\rm{loop}}=k$ for the $Q^{1,1,1}/\mathbf{Z}_{k}$~\cite{Aganagic:2009zk,Kim:2012vza} model, and by $\hat{Z}_{1-\rm{loop}}=2k$ for the  $Q^{1,1,1}/\mathbf{Z}_{k}$~\cite{Hanany:2008fj,Davey:2009sr,Franco:2009sp,Amariti:2011uw} and $M^{3,2}/\mathbf{Z}_{k}$~\cite{Martelli:2008si,Franco:2009sp} theories.

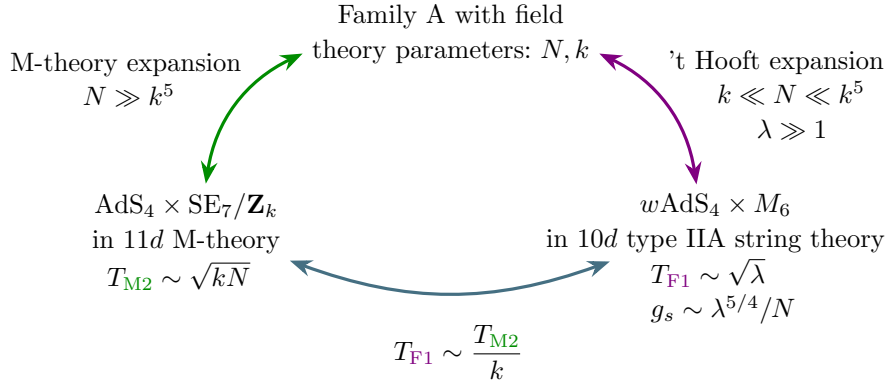
\begin{figure}[htbp]
	\centering
	
	\begin{tikzpicture}[
		>=Stealth,
		every node/.style={align=center},
		greenarrow/.style={<->, very thick, draw=green!55!black},
		purplearrow/.style={<->, very thick, draw=violet},
		bluearrow/.style={<->, very thick, draw=cyan!40!black},
		]
		
		% Nodes
		\node (top) at (0,2.8)
		{ Family A with field \\
			theory parameters: $N,k$};
		
		\node (left) at (-3.5,0.3)
		{ $\mathrm{AdS}_4\times \text{SE}_7/\mathbf Z_k$\\
			in 11$d$ M-theory};
		
		\node (right) at (3.5,0.3)
		{ $w\mathrm{AdS}_4\times M_6$ \\
			in 10$d$ type IIA string theory};
		
		% Arrows
		\draw[greenarrow]
		(left) to[bend left=28] (top);
		
		\draw[purplearrow]
		(top) to[bend left=28] (right);
		
		\draw[bluearrow]
		(right) to[bend left=20] (left);
		
		% Labels
		\node[align=right] at (-4.3,2.2)
		{M-theory expansion\\
			$N\gg k^5$\quad \phantom{aaa}};
		
		\node[align=left] at (-3.6,-0.4)
		{$T_{\text{\textcolor{green!55!black}{M2}}}\sim\sqrt{kN}$};
		
		\node[align=left] at (4.3,2)
		{'t Hooft expansion\\
			\phantom{aaa} $k\ll N\ll k^5$\\
			\phantom{aaaaaa} $\lambda\gg1$};
		
		\node[align=left] at (3.6,-0.6)
		{$T_{\text{\textcolor{violet}{F1}}}\sim\sqrt{\lambda}$\\
			$g_s\sim \lambda^{5/4}/N$};

		\node at (0.1,-1.4)
		{$T_{\text{\textcolor{violet}{F1}}}\sim\dfrac{T_{\text{\textcolor{green!55!black}{M2}}}}{k}$};
		
	\end{tikzpicture}
	
	\caption{Schematic gauge/gravity relation of family A.}
	\label{fig:gauge/gravity}
	
\end{figure}

To further elucidate this, we refer to Fig.~\ref{fig:gauge/gravity} which shows how the supergravity expansion of \eqref{WL expansion} should be performed when working in the M-theory or string theory limit.  From there, we observe that the partition function of a fundamental string should be expanded in the fixed but large dimensionless string tension $T_{\text{F1}}$ and small $g_{s}$ limit, yielding
\begin{gather}
	\log Z_{\text{F1}}=-S_{\text{cl}}+\log Z_{\text{1-loop}}+\mathcal{O}(1/k)+\mathcal{O}(1/T_{\text{F1}})\,.
\end{gather} 
Comparing this with the expansion of the M2-brane partition function
\begin{gather}
	\log Z_{\text{M2}}=-\hat{S}_{\text{cl}}+\log \hat{Z}_{\text{1-loop}}(k)+\mathcal{O}(1/T_{\text{M2}})\,,
\end{gather}
we find that  
\begin{equation}
	\hat{Z}_{\text{1-loop}}(k)=Z_{\text{1-loop}}+\mathcal{O}(1/k)\,.
\end{equation} 
That is, the one-loop M2-brane partition function contains all the perturbative corrections of $\mathcal{O}(1/k)$ of the string theory expansion. Hence, the fact that this object reduces to the one-loop string partition function in the case of $Q^{1,1,1}/\mathbf{Z}_{k}$, $Q^{2,2,2}/\mathbf{Z}_{k}$ and $M^{3,2}/\mathbf{Z}_{k}$ suggests that the string theory limit $k\gg 1$ of these theories might be exact.

Interestingly, a similar observation was recently made in Ref.~\citen{Tseytlin:2026rxv}. There an analogue of the holographic circular Wilson loop vev was studied in the AdS$_{3}\times S^{3}\times T^{4}\times S^{1}$ eleven-dimensional background, and the corresponding M2-brane partition function was evaluated at one-loop order. In this case, the one-loop M2-brane partition was also found to reduce to the one-loop string partition function, a result attributed to the (4,4) supersymmetry preserved by the mass spectrum of the quadratic fluctuations.~\cite{Tseytlin:2026rxv} However, our examples do not exhibit this type of supersymmetry, and we therefore currently lack a formal explanation for the apparent exactness of the string theory limit.

\section{Holographic Wilson loops in string theory}

We now switch to studying the holographic Wilson loop for the second class of three-dimensional SCFTs: family B. These theories  are given by three-dimensional U$(N)_{k_{1}}\times$U$(N)_{k_{2}}\times\dots \times$U$(N)_{k_{\mathcal{G}}}$ Chern-Simons-matter theories with all CS levels equal to $k$ and we define $n\coloneqq\mathcal{G}k$, where the 't Hooft coupling is given by $\tilde\lambda=N/n$. The matrix model for this family of SCFTs was studied in the large $N$ limit by Refs.~\citen{Jafferis:2011zi, Guarino:2015jca,Fluder:2015eoa} and the vev of the circular half-BPS Wilson loop located on the equator of $S^{3}$ at leading order was found to be~\cite{Fluder:2015eoa}
\begin{equation}\label{WL fam B LO}
	\text{Re}\log\langle W\rangle_{\text{B}}=\sqrt[3]{\frac{4\pi^{6}N}{\sqrt{3}\,n\,\text{vol}(\text{SE}_{5})}}\,.
\end{equation}
We highlight that in contrast to family A, there is no known example of family B in the literature for which this observable has been calculated up to next-to-leading order from a matrix model perspective. 
In the analysis that follows we use holography to provide a prediction for the subleading  correction to \eqref{WL fam B LO}, by studying the semiclassical quantization of a fundamental string in the dual supergravity backgrounds.

These backgrounds arise by taking the near horizon limit of $N$ D2-branes in massive type IIA string theory and are given by the following solution~\cite{Guarino:2015jca,Fluder:2015eoa}
\begin{equation}\label{metric string frame}
	\dd{s}^{2}=L^{2}\e^{\phi_{0}/2}\sqrt{5+\cos 2\al}\Big[\dd{s_{\text{AdS}_{4}}^{2}}+\frac{3}{2}\dd{\al}^{2}+\frac{6\sin^{2}\al}{3+\cos 2\al}\dd{s_{\text{KE}_{4}}^{2}}+\frac{9\sin^{2}\al}{5+\cos 2\al}\eta^{2}\Big]\,,
\end{equation}
with
\begin{equation}
		\e^{\Phi}=\e^{\phi_{0}}\frac{(5+\cos 2\al)^{3/4}}{3+\cos 2\al}\,,\qquad F_{0}=\frac{\e^{-5\phi_{0}/4}}{\sqrt{3}L}\,,
\end{equation}
and
\begin{equation}\label{fluxes mIIA}
	\begin{aligned}
			H_{3}&=\dd B_{2}=24\sqrt{2}L^{2}\e^{\phi_{0}/2}\frac{\sin^{3}\al}{(3+\cos2\al)^{2}}\omega\wedge\dd{\al}\,,\\
		F_{2}&=-L \e^{-3\phi_{0}/4}\sqrt{6}\Big(4\frac{\sin^{2}\al\cos\al}{(3+\cos2\al)(5+\cos2\al)}\omega \\ &\qquad\qquad\qquad\qquad+3\frac{3-\cos2\al}{(5+\cos2\al)^{2}}\sin\al\dd{\al}\wedge\eta\Big)\,,\\
		F_{4}&=L^{3} \e^{-\phi_{0}/4}\Big(6\,\text{vol}_{\text{AdS}_{4}}+12\sqrt{3}\frac{7+3\cos2\al}{(3+\cos2\al)^{2}}\sin^{4}\al\,\text{vol}_{\text{KE}_{4}}\\ &\quad\quad\quad\quad\quad+18\sqrt{3}\frac{(9+\cos2\al)\sin^{3}\al\cos\al}{(3+\cos2\al)(5+\cos2\al)}\omega\wedge\dd{\al}\wedge\eta\Big)\,,
	\end{aligned}
\end{equation}
where $\e^{\Phi}$ is the dilaton,  $F_{0}$ the Romans mass, $\omega$ is the  Kähler two-form of the four-dimensional Kähler-Einstein base manifold, whose metric is given by $\dd{s_{\text{KE}_{4}}^{2}}$ and $\eta$ is its contact one-form which satisfies $\dd{\eta}=2\omega$. For these backgrounds the supergravity limit is accessed when taking  $L\gg \ell_{s}$. We can relate this length scale with the field theory parameters $N,n$ by imposing the flux quantization condition, which yields\cite{Fluder:2015eoa} 
\begin{equation}\label{flux quant string}
	\Big(\frac{L}{\ell_{s}}\Big)^{6}=\frac{n^{1/4}\pi^{6}}{18^{7/8}}	\Big(\frac{N}{\text{vol}(\text{SE}_{5})}\Big)^{5/4}\,,\qquad  \e^{\phi_{0}}=\frac{2^{11/12}}{3^{1/6}n^{5/6}}\qty(\frac{\text{vol}(\text{SE}_{5})}{N})^{1/6}\,.
\end{equation}

As mentioned in the introduction, in contrast to the holographic duals of family A, we cannot uplift this solution to M-theory, as the dilaton never grows large enough in massive type IIA.~\cite{Aharony:2010af} We therefore turn to the study of the holographic Wilson loop via string theory, which is given by the partition function of a fundamental string with suitable boundary conditions. To access its one-loop correction we should expand the corresponding partition function around fixed but large dimensionless string tension $T_{\text{F1}}\sim\tilde{\lambda}^{1/3}$ 
and small string coupling $g_{s}$
\begin{equation}\label{WL expansion string}
	\langle W \rangle_{\text{B}}=Z_{\text{F1}} \approx \e^{-S_\text{cl}}Z_{1-\text{loop}}\,,
\end{equation}
where similarly as before, the classical action is the leading contribution evaluated on the classical minimal surface of the string.
Using the Green-Schwarz formalism, the classical bosonic string action in static gauge reads
\begin{gather}\label{string cl action}
	S_{\text{cl}}=\frac{1}{2\pi\ell_{s}^{2}}\int\text{vol}_{2}+\frac{i}{2\pi\ell_{s}^{2}}\int B_{2}\,,
\end{gather}
where $\text{vol}_{2}$ is the volume form over the induced string worldsheet and the last term contains the pull-back of the Kalb-Ramond $B$-field to the worldsheet. Our primary focus is the evaluation of the one-loop string partition function $Z_{1-\text{loop}}$ which as we shall see momentarily contains two contributions: one attributed to the quadratic fluctuations around the string's minimal surface, while the second stems from the coupling of the dilaton to the worldsheet. 

For completeness, we note that the classical configuration of the string corresponding to the above Wilson loop was found in Ref.~\citen{Fluder:2015eoa} to be given by the AdS$_{2}\subset\text{AdS}_{4}$ minimal surface at $\alpha=0$ with induced metric 
\begin{equation}\label{string ws metric}
	\dd{s}_{\text{F1}}^{2}=\frac{\ell^2}{\sinh^{2}\s}(\dd{\s}^{2}+\dd{\tau}^{2})\,,\qquad \ell^2 \coloneqq \sqrt{6}L^{2}e^{\phi_{0}/2}\,.
\end{equation}
Hence, the classical action reduces to the area of the worldsheet, resulting in~\cite{Fluder:2015eoa}
\begin{equation}\label{class action fam B}
	S_{\text{cl}}=-\sqrt[3]{\frac{4\pi^{6}N}{\sqrt{3}\,n\,\text{vol}(\text{SE}_{5})}}\,,
\end{equation}
and successfully matches the matrix model calculation \eqref{WL fam B LO}.

We can now continue with the evaluation of the one-loop string partition function, which receives a contribution from the Fradkin-Tseytlin (FT) action
\begin{equation}\label{FT action}
	S_{\text{FT}}=\frac{1}{4\pi}\int\text{vol}_{2}\Phi_{0} R^{(2)}+\frac{1}{2\pi}\int_{\partial}\dd{s} \Phi_{0} K\,,
\end{equation}
where $\Phi_{0}$ is the dilaton pulled back to the worldsheet, $R^{(2)}$ is the Ricci scalar of the string worldsheet and $K$ is the extrinsic curvature on its boundary.
It also includes a contribution from the quadratic fluctuations normal to the worldsheet, which are captured by the following action\cite{Cvetic:1999zs,Singh:2023olv}
\begin{gather}\label{one-loop string lagrangian}
\begin{aligned}
	S^{(2)}&=\frac{1}{2\pi\ell_{s}^{2}}\int \dd{\sigma}\dd{\tau}(\sqrt{\gamma}\mathcal{L}_{\text{bos}}+\mathcal{L}_{\text{ferm}})\,,\\
	\mathcal{L}_{\text{bos}}&=\frac{1}{2}(\tensor{D}{^{a}_{b}}\zeta^{b})^{2}-\frac{1}{2}\Big(\tensor{R}{^{i}_{aib}}+\tensor{K}{_{a}^{ij}}\tensor{K}{_{bij}}-\frac{i}{2}\epsilon^{ij}\nabla_{a}H_{bij}-\frac{1}{4}\tensor{H}{_{a}^{ci}}\tensor{H}{_{bci}}\Big)\zeta^{a}\zeta^{b}\,,\\
	\mathcal{L}_{\text{ferm}}&=-i\bar\theta \mathcal{P}^{ij}\Big\{\Gamma_{i}D_{j}-\frac{1}{8}\Gamma_{11}\Gamma_{i}^{\mu\nu}H_{j\mu\nu}+\frac{1}{8}F_{0} \e^{\Phi}\Gamma_{i}\Gamma_{j}\\
	&\qquad\qquad\quad-\frac{1}{8}\e^{\Phi}\qty(\Gamma_{11}\Gamma_{i}\slashed{F}_{2}\Gamma_{j}-\Gamma_{i}\slashed{F}_{4}\Gamma_{j})\Big\}\theta\,,
\end{aligned}
\end{gather}
with
\begin{gather}
	\tensor{D}{^{a}_{b}}\zeta^{b}=\nabla\zeta^{a}+ \tensor{A}{^{a}_{b}}\zeta^{b}\,,\quad \tensor{A}{_{i}^{ab}}=\tensor{\Omega}{_{i}^{ab}}+\frac{i}{2}\epsilon_{ij}\tensor{H}{^{jab}}\,,\quad \mathcal{P}^{ij}=\sqrt{\g}\g^{ij}-i\Gamma_{11}\epsilon^{ij}\,,
\end{gather}
where the eight scalar fluctuations are given by $\zeta^{a}$, with $a,b=1,\dots,8$, and the ten-dimensional Dirac spinors are given by $\theta$. Moreover, $\gamma$ is the determinant of the worldsheet metric \eqref{string ws metric}, $\Gamma_{i}$ are the ten-dimensional gamma matrices, and $\epsilon^{ij}$ is the two-dimensional Levi-Civita symbol.

Remarkably, due to the symmetries of the induced worldsheet and the ones of the ten-dimensional supergravity background the quadratic action simplifies significantly and reduces to 
\begin{equation}\label{F1 quad action}
	S^{(2)}= \frac{1}{2\pi\ell_{s}^{2}}\int\text{vol}_{2}(\zeta^{a}\mathcal {K}_{ab}\zeta^{b}+\bar\theta^{a}\mathcal{D}_{ab}\theta^{b})\,,
\end{equation}
with diagonal massive Klein-Gordon and Dirac quadratic operators on AdS$_{2}$, that contain a universal mass spectrum independent of the particular SE$_{5}$ manifold under consideration. By combining this with the FT action which evaluates to $S_{\text{FT}}=\log(6^{3/4}\e^{\phi_{0}}/4)$, the one-loop partition function is given by
\begin{equation}
Z_{1-\rm{loop}}=\frac{4\e^{-\phi_{0}}}{6^{3/4}}\sqrt{\frac{ \prod_{\text{f}}\det \mathcal{D}}{ \prod_{\text{b}}\det \mathcal{ K}}}\,.
\end{equation}

Similarly as in the case for the M2-brane, we may evaluate the functional determinants of the two-dimensional operators by using the Heat kernel on AdS$_{2}$. This requires some care, as the resulting effective action will suffer from a logarithmic UV divergence $\log{\Lambda}$ due to the introduction of the UV cut-off $\Lambda$ in the analysis. However, this divergence is universal, as its prefactor is given by the Euler characteristic of the worldsheet.~\cite{Giombi:2020mhz}  Therefore, it can be treated by the universal prescription of   Ref.~\citen{Giombi:2020mhz}, which instructs us that $\log{\Lambda}$ should be replaced by $-\log(2\pi\ell_{s}/\ell)$, where $\ell$ is the radius of AdS$_{2}$. We then find that the one-loop string partition function upon inserting \eqref{flux quant string} yields the following prediction for the vev of the Wilson loop for an infinite class of SCFTs belonging to family B up to subleading order~\cite{Gautason:2025bft}
\begin{gather}\label{WL fam B}
	\langle W\rangle_{\text{B}}=\Gamma\qty(\frac{2}{3})^{3}\sqrt[3]{\frac{n^{2}N}{12^{2}\pi^{3}\,\text{vol}(\text{SE}_{5})}}\exp\qty(\sqrt[3]{\frac{4\pi^{6}N}{\sqrt{3}\,n\,\text{vol}(\text{SE}_{5})}})\,.
\end{gather}

\section{Conclusions}

We studied the semiclassical quantization of a probe M2-brane/fundamental string in the dual geometry of family A/B in a unified manner, which led to novel predictions for the subleading correction to the vacuum expectation value of the half-BPS Wilson loop of both families. In particular, for family A we found the exact expression of the one-loop partition function of a probe M2-brane, in terms of the volume of the seven-dimensional Sasaki-Einstein manifold, the CS level $k$ and the rank of the gauge group $N$. 
In the context of the AdS$_{4}$/CFT$_{3}$ correspondence, this allowed us to perform a novel non-trivial holographic precision test of the duality between ADHM theory and M-theory on AdS$_{4}\times S^{7}/\mathbf{Z}_{N_{f}}$. On the other hand, our holographic answer for the perturbative part of the Wilson loop of family A matched the corresponding field theory result for the ABJM and ADHM theory, highlighting the predictive power of \eqref{WL pert} in combination with \eqref{eq: M2 eff action sum} and \eqref{eq: M2 eff action 2} for the remaining theories of family A.

Similarly to family A, we were able to express the one-loop string partition function dual to the Wilson loop of family B solely in terms of the volume of the five-dimensional Sasaki-Einstein manifold, the sum of the CS levels $n$ and the rank of the gauge group $N$. However, in this case the subleading correction to the Wilson loop vev is currently undetermined from the field theory point of view. Hence, \eqref{WL fam B} constitutes a prediction, but a non-trivial holographic precision test of family B and its dual theories remains to be established. This motivates a study of the associated matrix models beyond leading order.

In addition, several open questions remain, which warrant further investigation. First, we used the one-loop exactness of the M2-brane partition function to extract the exact perturbative answer for the Wilson loop, matching the field theory results for two examples of family A. However, it remains to be shown with an analogous analysis to the one performed here that the two-loop contribution to the partition function indeed vanishes. Second, given the recent proposals on holographic ensembles in M-theory~\cite{Gautason:2025plx} and type IIB string theory~\cite{vanMuiden:2026lno} it will be interesting to understand whether a similar proposal could hold for the case of massive type IIA. We will report on this in future work.~\cite{wip}
Finally, our results provide a promising framework for the direct investigation of a broader class of supersymmetric observables from their gravitational duals.

\section*{Acknowledgments}
It is a pleasure to thank Fridrik Freyr Gautason for the collaboration on the work underlying this proceedings contribution, for fruitful discussions, and for providing comments on the manuscript. I am very grateful to the organisers of the “Athens Workshop in Theoretical Physics: 10th Anniversary” for
the kind invitation to contribute to its proceedings. This work is supported by the Eimskip Fund of the University of Iceland.

\section*{ORCID}

\noindent Alexia Nix - \url{https://orcid.org/0009-0007-9920-3607}

\bibliographystyle{ws-ijmpd}
\bibliography{RefsWL}

\end{document}